\documentclass[useAMS,usenatbib,usegraphicx]{mn2e}

\newcommand{\kms}{km s$^{-1}$}

\newcommand{\lam}{$\lambda$}
\newcommand{\civ}{\mbox{C\,{\sc iv}}}

\newcommand{\siiv}{\mbox{Si\,{\sc iv}}}
\newcommand{\oiv}{\mbox{O\,{\sc iv}]}}

\title[Variability in Quasar Broad Absorption Line Outflows]
{Variability in Quasar Broad Absorption Line Outflows I. Trends in the Short-Term versus Long-Term Data}
\author[D. M. Capellupo, F. Hamann, J. C. Shields, P. Rodr\'iguez Hidalgo, T.A. Barlow]{D. M. Capellupo$^{1}$
\thanks{E-mail:dancaps@astro.ufl.edu (DMC)}, F. Hamann$^{1}$, J. C. Shields$^{2}$, P. Rodr\'iguez Hidalgo$^{3}$,\newauthor and T.A. Barlow$^{4}$\\
$^{1}$Department of Astronomy, University of Florida, Gainesville, FL 32611\\
$^{2}$Department of Physics \& Astronomy, Ohio University, Athens, OH 45701\\
$^{3}$Department of Astronomy and Astrophysics, Pennsylvania State University, University Park, PA 16802\\
$^{4}$Infrared Processing and Analysis Center, California Institute of Technology, Pasadena, CA 91125}
\begin{document}


\pagerange{\pageref{firstpage}--\pageref{lastpage}} \pubyear{2002}

\maketitle

\label{firstpage}

\begin{abstract}
Broad absorption lines (BALs) in quasar spectra identify high velocity outflows that likely exist in all quasars and could play a major role in feedback to galaxy evolution. The variability of BALs can help us understand the structure, evolution, and basic physical properties of the outflows. Here we report on our first results from an ongoing BAL monitoring campaign of a sample of 24 luminous quasars at redshifts 1.2$<$z$<$2.9, focusing on \civ\ \lam1549 BAL variability in two different time intervals: 4 to 9 months (short-term) and 3.8 to 7.7 years (long-term) in the quasar rest-frame. We find that 39\% (7/18) of the quasars varied in the short-term, whereas 65\% (15/23) varied in the long-term, with a larger typical change in strength in the long-term data. The variability occurs typically in only portions of the BAL troughs. The components at higher outflow velocities are more likely to vary than those at lower velocities, and weaker BALs are more likely to vary than stronger BALs. The fractional change in BAL strength correlates inversely with the strength of the BAL feature, but does not correlate with the outflow velocity. Both the short-term and long-term data indicate the same trends. The observed behavior is most readily understood as a result of the movement of clouds across the continuum source. If the crossing speeds do not exceed the local Keplerian velocity, then the observed short-term variations imply that the absorbers are $<$6 pc from the central quasar.
\end{abstract}


\section{Introduction}

Accretion disk outflows play a key role in the physics of quasars and their environs. Quasar
outflows may be an integral part of the accretion process and the growth of supermassive
black holes (SMBHs), by allowing the accreting material to release angular momentum. These
outflows may also provide enough kinetic energy feedback to have an effect on star formation
in the quasar host galaxies, to aid in `unveiling' dust-enshrouded quasars, and to help
distribute metal-rich gas to the intergalactic medium (e.g., \citealt{diMatteo},
\citealt{moll07}).

The location and three-dimensional structure of quasar outflows are poorly understood.
Sophisticated models have been developed that envision these outflows as arising from a
rotating accretion disk, with acceleration to high speeds by radiative and/or
magneto-centrifugal forces (\citealt{murray}, \citealt{proga04}, \citealt{proga07},
\citealt{everett}). Improved observational constraints are needed to
test these models and allow estimation of mass loss rates, kinetic energy yields, and the
role of quasar outflows in feedback to the surrounding environment. In this
paper, we focus on the most prominent signatures of accretion disk outflows that appear in
quasar spectra, the broad absorption lines (BALs). BALs are defined to have velocity widths
$>$2000 \kms\ at absorption depths $>$10\% below the continuum \citep{balnicity},
and they appear in the spectra of $\sim$10-15\% of quasars (\citealt{reichard03a},
\citealt{trump06}). Since only a fraction of quasar spectra display these features, the
presence of BALs could represent a phase in the evolution of a quasar and/or particular
orientations where the outflow lies between us and the quasar emission sources. Other
studies have found support for the latter scenario, given the similarity of various
properties between BALQSOs and non-BALQSOs (e.g., \citealt{balnicity}, \citealt{shen08}).

Studying the variability in these absorption features can provide important insight into the
structure and dynamics of the outflows. For example, we can use information about short-term
variability to place constraints on the distance of the absorbing material from the central
SMBH. The more quickly the absorption is varying, the closer the absorber is to the central
source, based on nominally shorter crossing times for moving clouds (\citealt{fred08}
and \S5 below) or the higher densities required for shorter recombination times
\citep{hamann97}. Long-term variability measurements provide information on the
homogeneity and stability of the outflow. A lack of variability on long time-scales implies
a smooth flow with a persistent structure. General variability results provide details on
the size, kinematics, and internal makeup of sub-structures within the flows. Variability
studies can also address the evolution of these outflows as the absorption lines are seen to
come and go, or one type of outflow feature evolves into another (e.g., \citealt{fred08}).

A small number of previous studies have investigated variability in BALs, emphasizing \civ\
$\lambda$1549 because of its prominence and ease of measurement. \citet{bar} carried out a
study of 23 BALQSOs, covering time-scales of $\Delta$t $\la$1 yr\footnote{Throughout this paper, all time intervals are measured in years in the rest frame of the quasar.}.
\citet{L07} used Sloan Digital Sky Survey (SDSS) spectra of 29 quasars to study \civ\ BAL
variability on similar time-scales. \citet{gibson08} studied variability on longer time-scales
(3-6 years) by combining data for 13 BALQSOs from the Large Bright Quasar Survey (LBQS) and SDSS. \citet{gibson10} reports on variability on multi-month to multi-year time-scales, using 3-4 epochs of data for 9 BALQSOs. They also compare variability in \siiv\ absorption to variability in \civ\ absorption in their sample.

In the present study, we go beyond existing work by carrying out a monitoring program of
BALQSOs from the sample of \citet{bar}. This strategy provides BAL variability data covering
longer time-scales as well as a wider range in time-scales within individual sources. We
include archival spectra from the SDSS, when available, to augment the temporal sampling.
This paper reports our results for \civ\ \lam1549 variability with measurements selected to
enable comparison between short-term (0.35 to 0.75 yr) and long-term (3.8 to 7.7 yr)
behavior. We identify trends in the data with velocity and the depth of the absorption.
We avoid using equivalent width (EW) measurements, which can minimize a change that occurs
in a portion of a much wider trough. In subsequent papers, we will discuss variability
properties in the entire data set, including more epochs on individual quasars, sampling in
a much shorter time domain, and comparisons between the \civ\ and \siiv\ $\lambda$1400
variabilities to place constraints on the outflow ionizations and column densities. Section
2 below discusses the observations and the quasar sample. Section 3 describes the steps we
followed to identify the \civ\ BALs and where they varied. Section 4 describes our results,
and Section 5 discusses the implications of these results with comparisons to previous work.

\section[]{Data and Quasar Sample}

Between 1988 and 1992, \citet{bar} and his collaborators (e.g., \citealt{barlow92}) obtained
spectra for 28 BALQSOs with the Lick Observatory 3-m Shane Telescope. They obtained
multi-epoch spectra for 23 of these quasars for a study of short-term BAL variability. They
selected these objects from already known BALQSOs, with redshifts of 1.2 to 2.9. When
selecting objects for their monitoring program, they gave preference to quasars known to
have either absorption line or photometric variability. However, in most cases, this
information was not available, and they selected more quasars at higher redshifts due to
higher detector sensitivity at longer wavelengths. While this sample of BALQSOs is not
randomly selected, it does cover a wide range of BAL strengths (see Table 1 and Figure 1).

\citet{bar} used the Lick Observatory Kast spectrograph to obtain spectra with settings for
high resolution, $R\equiv\lambda/\Delta\lambda\approx 1300$ (230\kms), and moderate
resolution, $R\approx 600$ (530 \kms). Most of the objects were observed at both settings.
For our analysis of these data discussed below, we used the high-resolution observations for
most sources. In some cases, only a moderate-resolution observation is available or choosing
the high-resolution observation would compromise wavelength coverage. Since BALs have a
width of at least 2000 \kms, a resolution of 530 \kms\ is sufficient to detect changes in
their profiles.

Starting in January 2007, we have been using the MDM Observatory 2.4-m Hiltner telescope to
reobserve the BALQSOs in the sample of \citet{bar}, with the goal of monitoring BAL
variability over a wide range of time-scales, from $<$1 month to nearly 8 years. We also
supplemented our data set with spectra from the Sloan Digital Sky Survey (SDSS). So far, we
have over 120 spectra for 24 BALQSOs, and we continue to collect more data. We note that two of these objects are not strictly BALQSOs because they have a balnicity index of zero (see \S3.2 below). Nonetheless, we include them in our sample because they do have broad absorption features at velocities that we include in our variability analysis (\S3.3).

We used the MDM CCDS spectrograph with a 350 groove per mm grating in first order and a
$1^{\prime\prime}$ slit to yield a spectral resolution of $R \approx 1200$ (250 km s$^{-1}$),
with wavelength coverage of $\sim 1600$ \AA . ÊThe spectrograph was rotated between
exposures to maintain approximate alignment of the slit with the parallactic angle, to
minimize wavelength-dependent slit losses. The wavelength range for each observation was
determined based on the redshift of the target quasar such that the coverage was blue enough
to include the Ly$\alpha$ emission and red enough to include the \civ\ emission feature. The
coadded exposure times were typically 1-2 hours per source.

We reduced the data using standard reduction techniques with the IRAF\footnote{The
Image Reduction and Analysis Facility (IRAF) is distributed by the National Optical Astronomy
Observatories (NOAO), which is operated by the Association of Universities for Research in
Astronomy (AURA), Inc., under cooperative agreement with the National Science Foundation.}
software. The data are flux calibrated on a relative scale to provide accurate spectral
shapes. Absolute flux calibrations were generally not obtained due to weather or time
constraints.

The Sloan Digital Sky Survey is an imaging and spectroscopic survey of the sky at optical
wavelengths. The data were acquired by a dedicated 2.5-m telescope at Apache Point
Observatory in New Mexico. The spectra cover the observer-frame optical and near-infrared,
from 3800 to 9200 \AA, and the resolution is $R\approx 2000$ (150 \kms). For spectra,
typically three exposures were taken for 15 minutes each, and more exposures were taken in
poor conditions to achieve a target signal-to-noise ratio. The multiple observations were
then co-added \citep{sloan}. The SDSS includes
observations of 11 of the quasars in the Lick sample. However, 3 of the SDSS observations
are unusable because the redshifts of those quasars are too low for the \civ\ BALs to appear
in the SDSS spectra. Therefore, we have usable SDSS spectra for 8 of the objects in our
sample, and they were taken between 2000 and 2006.

\section{Analysis}

\begin{table*}
  \begin{minipage}{185mm}
    \caption{Quasar Data and CIV Variability Results}
    \begin{tabular}{clccllccllcc}
\hline
Coord. & & & & \multicolumn{4}{c}{--------------- Short-Term ---------------} & \multicolumn{4}{c}{--------------- Long-Term ---------------} \\
(1950.0)& Name & $z_{em}$ & BI & Obs. 1 & Obs. 2 & $\Delta$t & Vary? & Obs. 1 & Obs. 2 & $\Delta$t & Vary? \\
\hline
0019+0107   & UM 232            & 2.13 & 2290  & 1989.84 & 1991.86 & 0.65 & N & 1989.84 & 2007.04 & 5.50 & N \\
0043+0048   & UM 275            & 2.14 & 4330  & 2000.69 & 2001.79 & 0.35 & N & 1991.86 & 2008.03 & 5.15 & Y \\
0119+0310   & AD85 D08          & 2.09 & 5170  & 1989.85$^{a}$ & 1991.86 & 0.65 & Y & 1989.85$^{a}$ & 2007.07 & 5.57 & Y \\
0146+0142   & UM 141            & 2.91 & 5780  & 1988.93$^{a}$ & 1991.86 & 0.75 & Y & 1988.93$^{a}$ & 2007.05 & 4.64 & Y \\
0226$-$1024 & WFM91 0226$-$1024 & 2.25 & 7770  &         &         &      &   & 1991.87 & 2007.04 & 4.66 & Y \\
0302+1705   & HB89 0302+170     & 2.89 & 0     &         &         &      &   & 1989.84$^{a}$ & 2007.05 & 4.42 & N \\
0842+3431   & CSO 203           & 2.15 & 4430  & 2007.04 & 2008.35 & 0.42 & Y & 1990.90 & 2008.35 & 5.54 & N \\
0846+1540   & H 0846+1540       & 2.93 & 0     & 1990.16 & 1992.19 & 0.52 & Y & 1992.19 & 2007.05 & 3.78 & Y \\
0903+1734   & HB89 0903+175     & 2.77 & 10700 & 2006.31 & 2008.28 & 0.52 & Y & 1989.26$^{a}$ & 2008.28 & 5.04 & Y \\
0932+5006   & SBS 0932+501      & 1.93 & 7920  & 1989.84 & 1991.10 & 0.43 & N & 1989.84 & 2008.28 & 6.30 & Y \\
0946+3009   & PG 0946+301       & 1.22 & 5550  & 1991.10 & 1992.23 & 0.51 & Y & 1991.10 & 2008.28 & 7.74 & Y \\
0957$-$0535 & HB89 0957$-$055   & 1.81 & 2670  & 1990.89 & 1992.32 & 0.51 & N & 1992.32 & 2008.35 & 5.70 & N \\
1011+0906   & HB89 1011+091     & 2.27 & 6100  & 2007.07 & 2008.35 & 0.39 & N & 1992.19 & 2008.35 & 4.94 & Y \\
1232+1325   & HB89 1232+134     & 2.36 & 11000 &         &         &      &   & 1989.26$^{a}$ & 2008.03 & 5.58 & N \\
1246$-$0542 & HB89 1246-057     & 2.24 & 4810  & 2007.04 & 2008.35 & 0.40 & N & 1992.19 & 2008.35 & 4.99 & Y \\
1303+3048   & HB89 1303+308     & 1.77 & 1390  & 2007.04 & 2008.28 & 0.45 & N & 1992.32 & 2008.28 & 5.76 & Y \\
1309$-$0536 & HB89 1309-056     & 2.22 & 4690  &         &         &      &   & 1992.19 & 2007.04 & 4.61 & N \\
1331$-$0108 & UM 587            & 1.88 & 10400 & 2007.04 & 2008.28 & 0.43 & N & 1992.32 & 2008.28 & 5.55 & Y \\
1336+1335   & HB89 1336+135     & 2.45 & 7120  &         &         &      &   & 1989.26$^{a}$ & 2008.28 & 5.52 & N \\
1413+1143   & HB89 1413+117     & 2.56 & 6810  & 2006.31 & 2008.28 & 0.55 & Y & 1989.26$^{a}$ & 2008.28 & 5.35 & Y \\
1423+5000   & CSO 646           & 2.25 & 3060  & 2007.07 & 2008.35 & 0.39 & N & 1992.32 & 2008.35 & 4.93 & N \\
1435+5005   & CSO 673           & 1.59 & 11500 &         &         &      &   & 1992.19 & 2008.35 & 6.25 & Y \\
1524+5147   & CSO 755           & 2.88 & 2490  & 1989.60 & 1991.52 & 0.49 & N & 1991.52 & 2008.28 & 4.32 & Y \\
2225$-$0534 & HB89 2225-055     & 1.98 & 7920  & 1988.46 & 1989.84 & 0.46 & N &         &         &      &   \\
\hline
    \end{tabular}
    \footnotetext[1]{ These Lick observations were taken at the lower resolution setting
      ($R\approx 600$; see \S2).}
  \end{minipage}
\end{table*}

\subsection{Short-Term and Long-Term Data Sets}

From the data in our BAL monitoring campaign, for each object, we select one pair of
observations for the short-term analysis and one pair for the long-term analysis. For the
long-term analysis, we want the longest time baseline possible. The first observation for
each object is the earliest Lick observation with the highest available resolution as well
as wavelength coverage that extends from the Ly$\alpha$ emission line through the \civ\
emission line. For the second observation for each object, we use the most recent MDM
spectrum. We have long-term data for 23 objects, with resulting $\Delta$t of 3.8 to 7.7
years. We note that for certain objects, the redshift is low enough that the Ly$\alpha$
emission line is not shifted into the wavelength coverage of our data and therefore is not
present in some of the spectra (e.g., 0946+3009).

For the short-term analysis, the range in the values of $\Delta$t among the objects should
be small, while including as many objects as possible. The best compromise we found was a
range for $\Delta$t of 0.35 to 0.75 yr, which allows us to include 18 objects in the
short-term analysis. In order to include this many objects, we use data for some objects 
that only covers the region from the \siiv\ emission to the \civ\ emission, which is sufficient for the analysis in this paper. These observations were selected separately from the long-term epochs and were chosen to fit into this $\Delta$t interval. These observations were taken before
or at any time in between the two long-term observations. There is one object for which we
have short-term Lick data, but no long-term data (2225-0534), so we have in total 24 objects
in our full sample.

\subsection{Characterizing the Quasar Sample}

Table 1 summarizes our full sample of long-term and short-term data, as described above. The first four columns provide information on all 24 quasars. $z_{em}$ is the emission redshift\footnote{The values of $z_{em}$ were obtained from the NASA/IPAC Extragalactic Database (NED), which is operated by the Jet Propulsion Laboratory, California Institute of Technology, under contract with the National Aeronautics and Space Administration.},
and BI is the ``balnicity index,'' as described below. The remaining columns are separated
into the short-term analysis and long-term analysis. The columns with ``vary?'' have either
Yes or No values to indicate variability anywhere in the \civ\ troughs (refer to \S4), and
$\Delta$t is the time difference (in years) in the quasar frame between observations.
The observations from 1988 to 1992 are from Lick, the observations from 2000 to 2006 are
from SDSS, and the observations from 2007 to 2008 are from MDM.

\begin{figure*}
  \includegraphics[width=160mm]{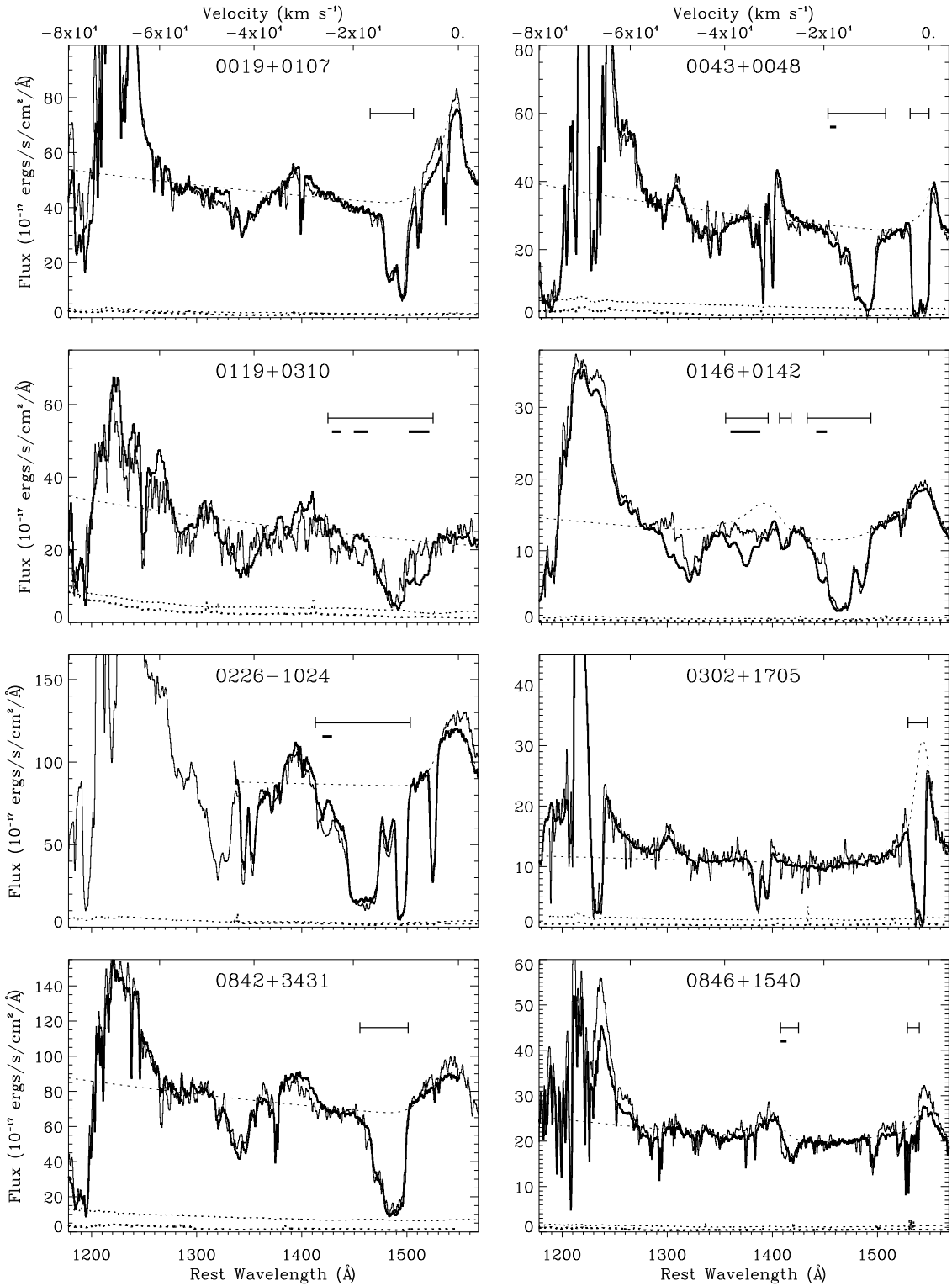}
  \caption{Smoothed spectra of all 24 quasars in our sample, showing the long-term
    comparisons between a Lick Observatory spectrum (bold curves) and the most recent MDM
    data (thin curves). The one exception is 2225-0534, for which we only have short-term
    data (see Table 1). The vertical flux scale applies to the Lick data, and the MDM
    spectrum has been scaled to match the Lick data in the continuum. The dashed curves show
    our pseudo-continuum fits. The thin horizontal bars indicate intervals of BAL absorption
    included in this study, and the thick horizontal bars indicate intervals of variation
    within the BALs. The formal 1$\sigma$ errors are shown near the bottom of each panel.}
\end{figure*}
\addtocounter{figure}{-1}
\begin{figure*}
  \includegraphics[width=170mm]{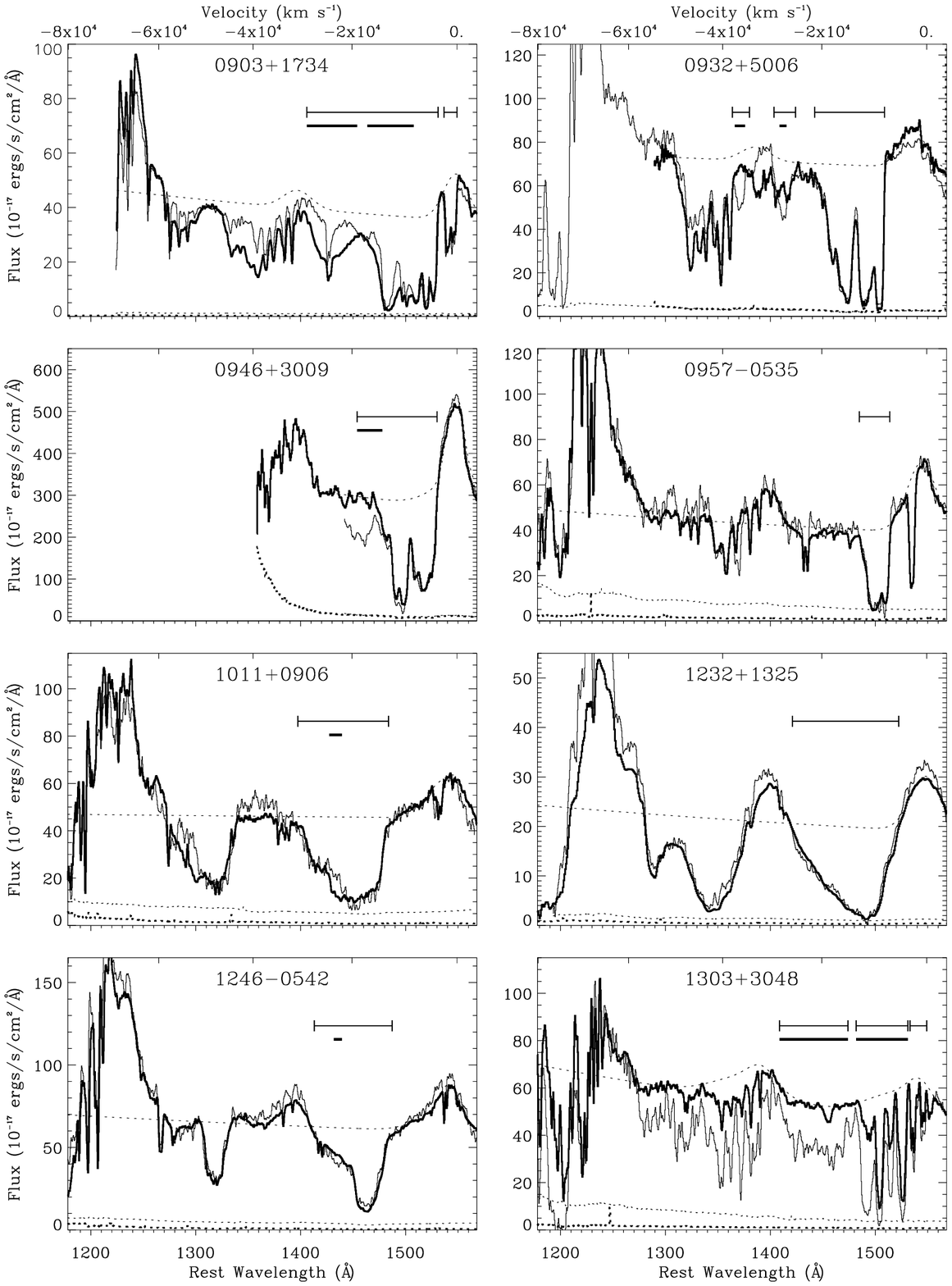}
  \caption{continued...}
\end{figure*}
\addtocounter{figure}{-1}
\begin{figure*}
  \includegraphics[width=170mm]{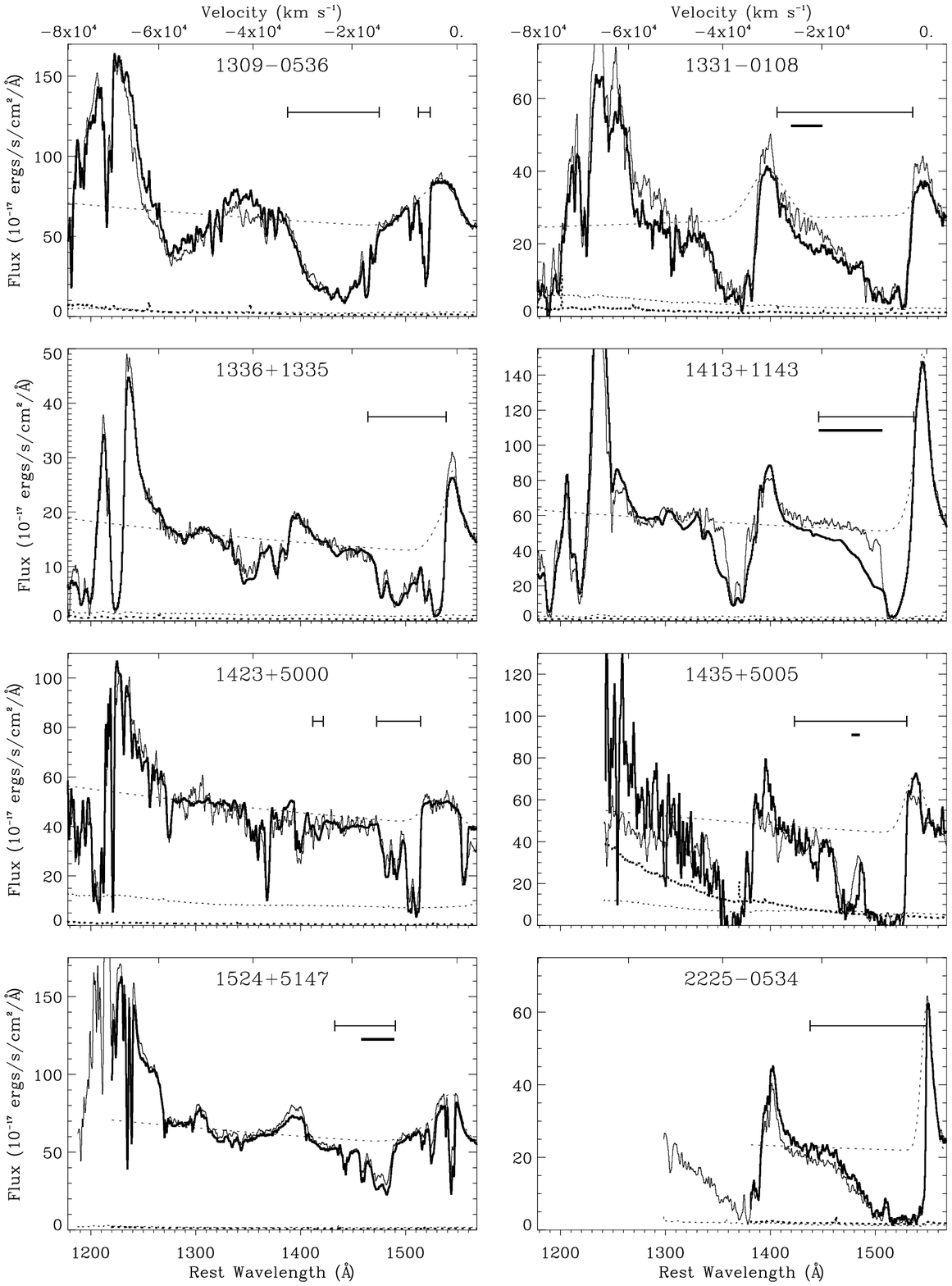}
  \caption{continued...}
\end{figure*}
Figure 1 contains spectra for all 24 objects, showing the long-term comparisons between a
Lick spectrum and an MDM spectrum. The one exception is 2225-0534, for which we only have
short-term Lick data. The velocity scale is based on the wavelength of \civ\ in the observed
frame calculated from the redshifts given in Table 1. We use a weighted average of the
\civ\ doublet wavelengths, i.e., 1549.05 \AA.

The balnicity index (BI, \citealt{balnicity}) is designed to differentiate between BALQSOs
and non-BALQSOs and provide a measure of the BAL strength, expressed as an equivalent width in units of velocity. It quantifies absorption that is blueshifted between $-$3000 and
$-$25000 \kms\ from the \civ\ emission line centre, is at least $\sim$10\% below the
continuum, and is wider than 2000 \kms. By this definition, if a quasar has a BI greater
than zero, then it is classified as a BALQSO. The highest possible value of BI is 20,000.

We measure BI in one observation per object to characterize the sample and allow for comparison with BALQSOs in other studies. We do not use BI in our variability analysis. In order to normalize the spectra and measure BI, and later absorption strength (see \S3.3), we first had to fit a pseudo-continuum to one spectrum per quasar, which includes the quasar's continuum emission plus the broad emission lines. We adopt the Lick observations used in the long-term analysis as our fiducial epoch for which we fit the pseudo-continua and measure BI. We constructed this pseudo-continuum by fitting a power-law to the continuum away from any emission or absorption lines, and then adding to that an estimate of the \civ\ and, if necessary, the \siiv\ broad emission line (BEL) profiles\footnote{Note that the feature we attribute to \siiv\ may also include \oiv\ \lam1400 emission.}. To fit the continuum, we fit a power-law function to two regions on the spectrum that are free of strong emission lines. The preferred wavelength ranges for the fits were 1270 to 1350 \AA\ and 1680 to 1800 \AA. These ranges were adjusted to avoid emission and absorption features as much as possible or due to the limits of the wavelength coverage. Fitting the \civ\ emission is complicated by, for example, potential asymmetries in the emission profiles and absorption in the emission region (e.g., \citealt{reichard03b}). Therefore, we are flexible in our fitting routine, using between 1 and 3 gaussians to define the line profile. The gaussians have no physical meaning, and we simply use them for obtaining a reasonable fit to the profile. If there is any absorption on top of the emission, we ignore those wavelengths when defining the fit. In cases where the absorption is wide enough that it obscures up to half of the emission line (e.g. 1413+1143), we assume the emission line is symmetric. In cases where the gaussian, or multiple-gaussian, fit did not provide the right shape and we have enough information about the profile (e.g. 1423+5000), we manually adjusted the fit to better match the \civ\ emission line.

For quasars where the \civ\ absorption is at a velocity that overlaps the \siiv\ emission,
we also fit the \siiv\ emission line.  If most, or all, of the \siiv\ emission feature is
absorbed, we synthesized a \siiv\ profile by taking the \civ\ fit, shifting its centre
wavelength to the location of \siiv, adjusting the FWHM (given the wider separation between
the \siiv\ doublet lines), and scaling the emission strength by a factor of 0.342. This
scale factor is based on composite quasar spectra from which \citet{vanden01} found the
typical strength of various emission lines. We relied on this scale factor in cases such as
0146+0142, where we have little information about the \siiv\ emission line. Otherwise, in a
case such as 0846+1540, we fit between 1 and 3 gaussians to the \siiv\ line, as we did for
the \civ\ emission, because we have more information about the profile of the \siiv\
emission line. The pseudo-continuum fits for each object are shown by the smooth dotted
curves in Figure 1.

\subsection{Measuring the BAL Variability}

For each quasar, we compare the two spectra in our short-term and long-term intervals (see
Table 1) to search for variability in the \civ\ BALs. First, we determined the velocity
ranges over which BAL absorption occurs. These ranges might differ between the two observing
epochs. We therefore set the range to the extremes observed in either epoch. Our definition
of these ranges is guided by the definition of BI, i.e. they must present contiguous
absorption reaching $\geq$10\% below the continuum across $\geq$2000 \kms. We do not follow
the requirement that the absorption be blueshifted between $-$3000 and $-$25000 \kms\
because we want to include all \civ\ broad absorption in our analysis (as a result, our
sample contains two objects, 0302+1705 and 0846+1540, with a BI of 0). In general, the \civ\
BALs occur between the \siiv\ and \civ\ emission lines, but some of the \civ\ absorption is
blue-shifted enough that it appears blueward of the \siiv\ emission line center (e.g., 0146+0142). In these cases, we can confirm that the absorption blueward of the \siiv\ emission line is due to \civ, and not \siiv, because any significant \siiv\ absorption should be accompanied by \civ\ absorption at the same velocity. This is evident from numerous observations of BALQSOs where the \siiv\ and \civ\ absorption are clearly identified (e.g., Figure 1, Korista et al. 1993), and from theoretical considerations of the relative strengths of these lines (Hamann et al. 2008). Our sample includes broad absorption troughs from 0 to $-$40000 \kms.

In order to compare any two spectra for variability in the defined absorption ranges, we scaled the spectra so that regions free of absorption and emission in each spectrum match. This is necessary because of uncertainties in the flux calibration and possible real changes in the quasar emissions. We applied a simple vertical scaling so that the spectra from each epoch match the fiducial Lick spectrum along the continuum redward of the \civ\ emission feature (i.e., from 1560 \AA\ to the limit of the wavelength coverage), between the \siiv\ and \civ\ emission ($\sim$1425-1515 \AA), and between the Ly$\alpha$ and the \siiv\ emission ($\sim$1305-1315 \AA). In nearly all cases, a simple scaling produced a good match between the spectra from different epochs. The results for the two long-term epochs are shown in Figure 1. In some cases, however, there were disparities in the overall spectral shape of the two spectra. To remove these disparities, we fit either a linear function (for 0903+1734, 1413+1143, and 1423+5000) or quadratic function (for 0019+0107 and 1309-0536) to the spectral regions that avoid the BALs in a ratio of the two spectra. We then multiplied this function by the comparison spectrum.

We identify BAL variability in velocity intervals that correspond to at least 2 resolution elements in the lower resolution Lick spectra, i.e., 1200 \kms. Our procedure to identify variability intervals relies mainly on visual inspection because photon statistics alone are not sufficient for defining real variability. Most of the occurrences of variability recorded for this study are obvious by simple inspection. However, we also examined all cases of weak or possible variability to ensure that no occurrences of significant real variability are missed. We determined the significance of variability based on two criteria. First, we calculate the average flux and associated error within each variability interval and use that to place an error on the flux differences between the two spectra. We include in this study only the intervals of variability where the flux differences are at least 4-sigma. We emphasize that all of the intervals which varied by 4-sigma or more are readily identified by our initial inspection technique. Therefore, there is no danger of variable intervals being missed. However, the 4-sigma threshold alone can include spurious or highly uncertain variability results because of uncertainties not captured by the photon statistics. These additional uncertainties can arise from the flux calibration, a poorly constrained continuum placement, underlying emission line variability or our reconstruction of a severely absorbed emission line profile. We therefore apply a second significance criterion based on our assessment of these additional uncertainties. We take a conservative approach by excluding variability intervals where the additional uncertainties might be large (even if the flux difference passes the formal 4-sigma threshold described above). This second criterion caused the exclusion of just a few intervals. See below for examples. The velocity intervals finally classified as variable in the long-term data are indicated by bold horizontal lines in Figure 1.

To help clarify our assessments of the additional uncertainties, we point out some specific velocity intervals that pass the 4-sigma threshold and appear variable by casual inspection, but are nonetheless excluded from our analysis because the additional uncertainties are significant. For example, in 1246-0542, we do not include the region around $-$17,000 \kms\ because the noise level in that portion of the trough makes it a very tenuous candidate for inclusion in the study. The region around $-$21,000 \kms\ in 1435+5005 is a similar case because here the increased noise level of the spectra towards greater outflow velocities makes a determination of variability more uncertain. We note that the increased scatter in the data for these two objects is less noticeable in the smoothed spectra plotted in Figure 1. In a couple cases, a slight difference in the slope between two comparison spectra adds some uncertainty. In 1413+1143, the blue wing of the \civ\ trough in the MDM spectrum extends slightly above the defined continuum. This may be due to some anomaly in the flux calibration, so we define the blue end of the BAL (and the variability interval) as $-$20,600 \kms. In 2225-0534, there is very little continuum available for matching the two comparison spectra, and this added uncertainty makes the small flux difference at the bluest portion of the BAL trough too tenuous to be included. A slight difference in slope between the two epochs could be responsible for any apparent changes in the wing of this BAL, and we do not have enough visible continuum to discount this possibility.

Identifying variability in the spectral regions which overlap the \civ\ and \siiv\ emission lines is further complicated by possible variability in the emission lines themselves (e.g., \citealt{wilhite05}). We record BAL variability in these spectral regions only if it is large compared to any reasonable change in the broad emission lines and/or the absorption changes abruptly over just portions of a BEL profile. For example, in 0146+0142, the interval from $-$38,800 to $-$32,700 \kms\ clearly shows BAL absorption weakening to approximately continuum level. In another case, 0932+5006, we include the intervals from $-$37,500 to $-$35,400 \kms\ and $-$28,400 to $-$27,000 \kms\ because there is clearly BAL absorption there in the MDM spectrum. We did not attribute the change in flux from approximately $-$34,000 to $-$30,000 \kms\ to a BAL because it is unclear what is the primary cause of variability in this interval. While the identifications of BAL absorption and variability in the regions overlapping \civ\ and \siiv\ emission are secure, the measurements of absorption strength ($\langle A\rangle$; see below) in this region have added uncertainty since fitting the emission lines is more uncertain than fitting the continuum. When looking at correlations with velocity in this study, we identify trends using all the data, but also confirm the existence of these trends in the velocity range $-$27,000 to $-$8,000 \kms, which avoids the emission lines. For correlations with absorption strength (see below), since there are few cases of variability at the extremes of the examined velocity range, their inclusion does not significantly alter the plots or the final results.

\begin{figure}
  \includegraphics[width=84mm]{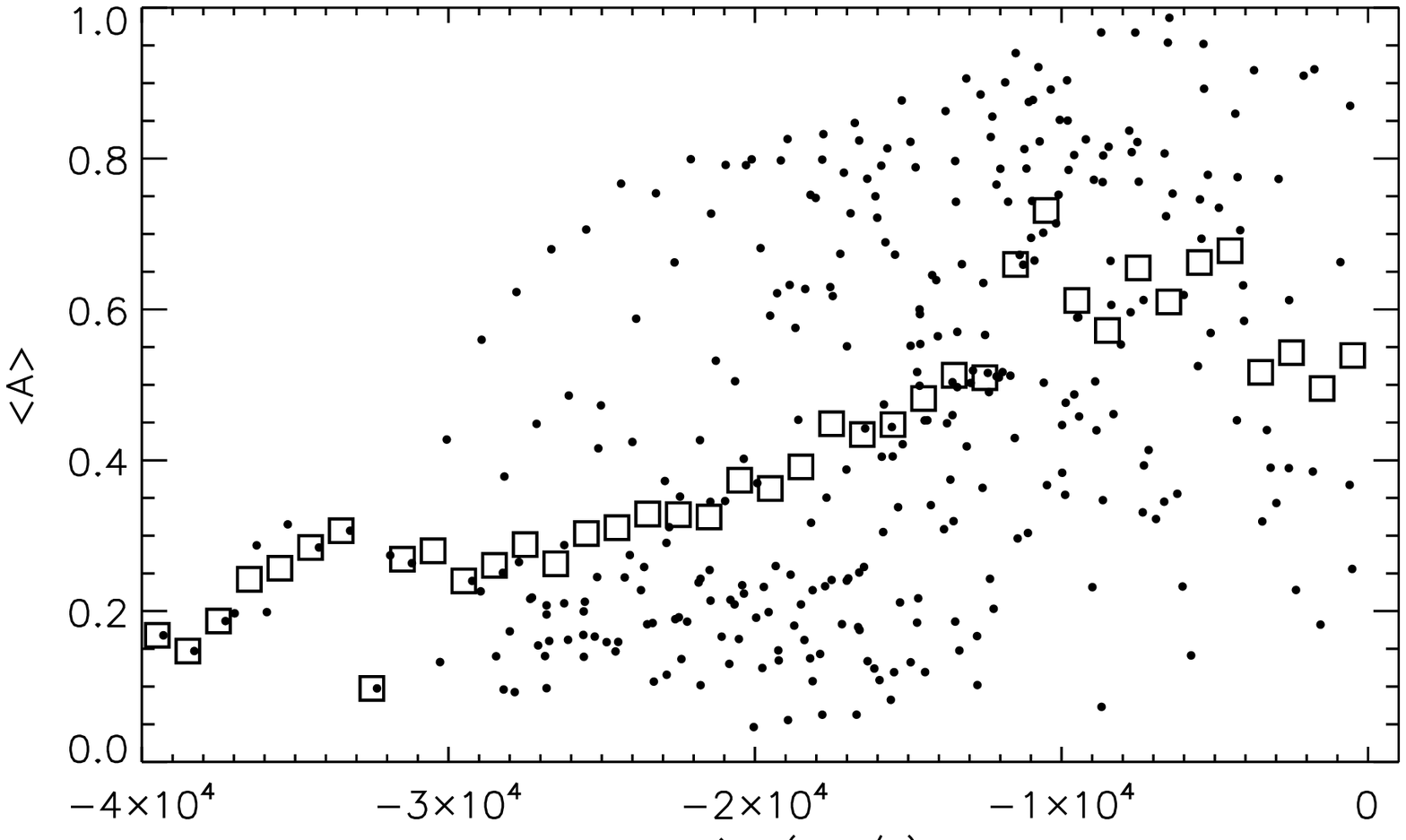}
  \caption{The average normalized absorption strength ($\langle A\rangle$), calculated throughout
    all of the BALs in the quasars in our long-term subsample, as a function of the outflow
    velocity of the absorbing material. The individual points in each bin (with width 1000
    \kms) represent the $\langle A\rangle$ for a single quasar, and the open squares represent the
    average of all points in that bin.}
\end{figure}
In order to look for trends in the data, we calculated the absorption strength as a function of velocity in each spectrum. We first normalized each spectrum using the pseudo-continuum fit (see \S3.2) for that quasar. We can apply the pseudo-continua derived in \S3.2 to the other epochs for each quasar because those epochs have already been scaled to match the fiducial Lick spectrum. However, we adjust the emission line fits in cases where the emission line varied. We define absorption strength, $A$, as the fraction of the normalized continuum flux removed by absorption (0 $\leq A\leq$ 1) within a specified velocity interval. A single $A$ value is adopted for each velocity interval based on the average flux within that interval. If there is variability, then the absorption strength changed and we calculate the average absorption strength, $\langle A\rangle$, and the change in absorption strength, $\Delta A$, between the two epochs being compared. The velocity intervals used to calculate $\langle A\rangle$ and $\Delta A$ are defined to i) include only portions of the spectrum with BAL absorption and ii) clearly separate the spectral regions within BAL troughs that did and did not vary. Specifically, the variable and non-variable absorption regions are each divided into equal-sized bins, with a bin width of 1000 to 2000 \kms\ depending on the velocity width of the region. This strategy makes the bin widths used to measure $\langle A\rangle$ and $\Delta A$ similar to the nominal width of 1200 \kms\ used above, while providing the flexibility needed to begin and end at the boundaries defined above between absorbed/unabsorbed and variable/non-variable spectral regions. In each of these velocity bins, we also calculate the fractional change in the absorption strength, $|\Delta A|/\langle A\rangle$, which can have values from 0 to 2. Spectral regions that did not vary according to the criteria described above have $\Delta A$ set to zero automatically. Note that, with $\langle A\rangle$ and $\Delta A$ determined in this way, each quasar can contribute more than once to these quantities in our variability analysis, depending on the range of strengths and velocities covered by its BAL troughs, because each bin showing variability is treated as a separate occurrence.

We plot the relationship between absorption strength, $\langle A\rangle$, and outflow velocity for all the quasars in our long-term subsample in Figure 2. A correlation between these two quantities will have implications for our results when we compare variability measures to each of these quantities individually (see \S4 and \S5). There is a trend for lower values of $\langle A\rangle$ at increasing outflow velocity, which is especially apparent from $-$27,000 to $-$8,000 \kms, away from the regions of \siiv\ and \civ\ emission. The errors in the individual points are much smaller ($\sigma_{\langle A\rangle}$$\le$0.01) than the dispersion of points in each bin. However, the bins at $<$-30,000 \kms\ only contain one or two quasars, so the average $\langle A\rangle$ in those bins can be uncertain by up to a factor of 2, given the additional uncertainty of fitting the \siiv\ emission line. The weaker absorption found at higher velocities in our sample is consistent with absorption in the mean BAL spectra constructed by \citet{korista93}.

\section{Results}

Table 1 indicates whether there was significant variability in the \civ\ BALs of each
quasar in the short-term and long-term data. We found that 39\% (7/18) of quasars with
short-term data varied, whereas in the long-term, 65\% (15/23) varied. One quasar varied
only in the short-term, so combining both short-term and long-term data, 67\% (16/24) of the
quasars varied overall.

\begin{figure}
  \includegraphics[width=84mm]{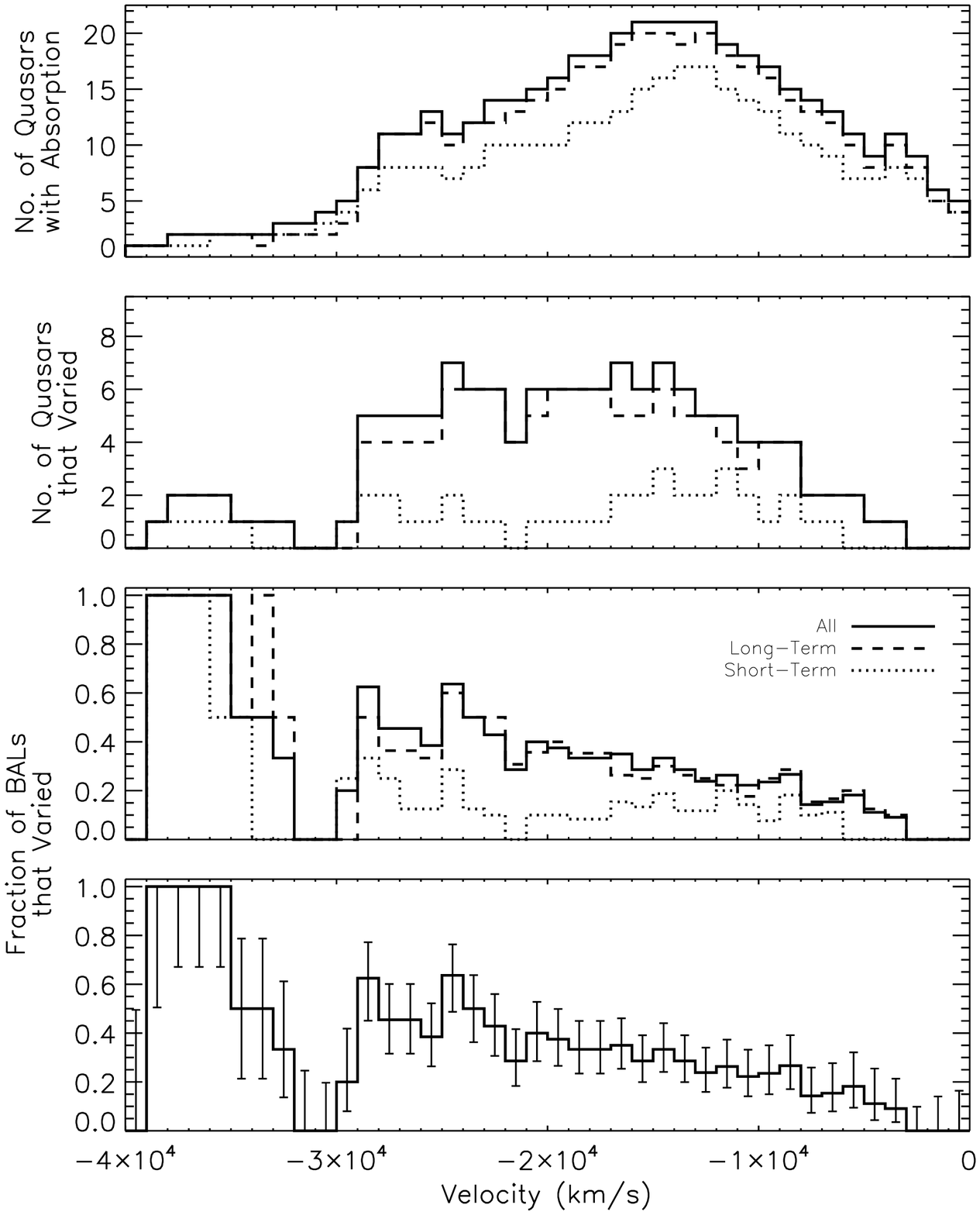}
  \caption{The top two panels show the number of quasars with \civ\ BAL absorption and the
    number with \civ\ BAL variability at each velocity. The third panel is the second panel divided by
    the top one, which gives the fraction of \civ\ BALs that varied at each velocity. The dashed, dotted,
    and solid curves indicate results for the long-term, short-term, and combined data set
    comparisons, respectively. The bottom panel displays the combined data set with 1$\sigma$
    error bars.}
\end{figure}
We investigated \civ\ BAL variability as a function of outflow velocity. Using the velocity
ranges for each quasar where we have identified absorption and variability (\S3.3), we count
how many quasars have BAL absorption and variability in each velocity bin. In Figure 3, the
top two panels show the number of quasars with \civ\ BAL absorption and the number with
\civ\ BAL variability at each velocity. The third panel is the second panel divided by the top one, which gives the fraction of \civ\ BALs that varied at each velocity. The dashed, dotted and solid curves indicate results for the long-term, short-term and combined data set comparisons, respectively. The plot implies that BALs at higher velocities are more likely to vary than those at lower velocities.

The bottom panel in Figure 3 displays 1$\sigma$ error bars based on \citet{wilson1927} and \citet{agresti98}. These errors are based on counting statistics for the number of quasars with absorption and variability at each velocity. We perform a least-squares fit to confirm that a correlation exists between incidence of variability and velocity. The slope of the combined data is $-1.51 \pm 0.25 \times 10^{-5}$, in the formal unit, fraction per \kms. The slope is non-zero at a 6-sigma significance. If we repeat the least-squares fit with only the bins with the most data (bins containing at least 8 quasars with absorption), the slope is $-1.76 \pm 0.33 \times 10^{-5}$ fraction per \kms, which is non-zero at a 5-sigma significance. Using this restriction limits the fit to only the data from $-$29,000 to $-$2,000 \kms. We perform the least-squares fit a third time, where we restrict the fit to only the data in the range $-$27,000 to $-$8,000 \kms\ that completely avoids overlap with the \civ\ and \siiv\ broad emission lines. In this case, the resulting slope is $-1.46 \pm 0.54 \times 10^{-5}$ fraction per \kms, which is non-zero at a 2.7-sigma significance. Therefore, this trend is evident even when we remove the data at the lowest and highest velocities, where additional uncertainty from the BELs can affect the results.

An important question is whether the trend seen in Figure 3 is a correlation with velocity in an absolute sense, or the cumulative result of differences between the red versus blue-side behaviors in individual absorption troughs. To address this question, we divided the velocity
range of absorption for each quasar into a red and a blue half and plotted the incidence of
variability separately for the red and blue portions. The result in both cases is consistent
with the behavior shown in Figure 3. We do not see a higher incidence of variability in the blue portions of troughs versus the red portions. This indicates that the trend in Figure 3 is a trend in the ensemble dependent on the absolute velocity rather than the relative velocity within a given trough. Consistent with this finding is the fact that none of the BALs varied at a velocity between $-$3500 \kms\ and 0 \kms, while the two BALs in our sample at $<$$-$33,000 \kms\ both varied.

\begin{figure}
  \includegraphics[width=84mm]{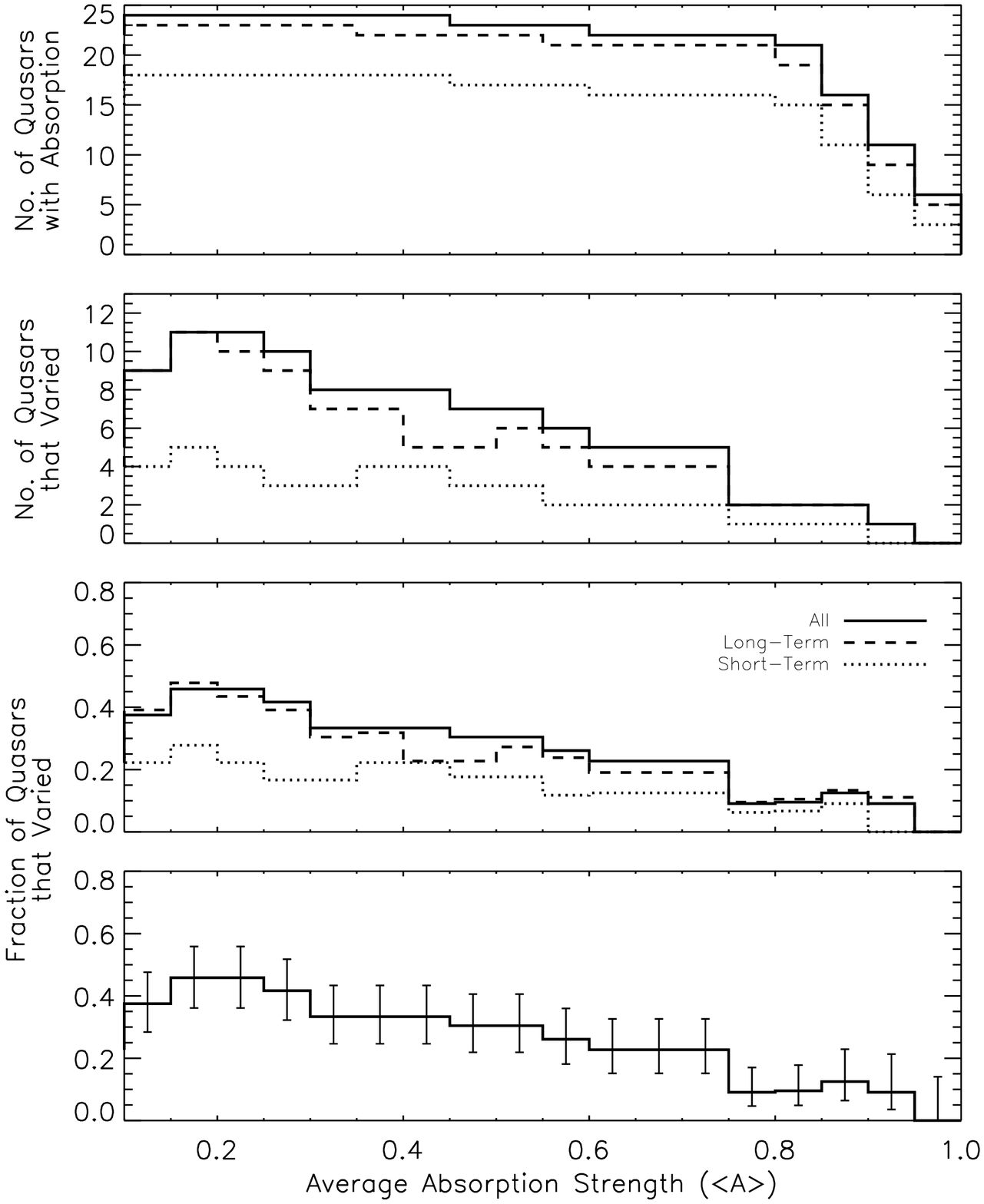}
  \caption{Same as Figure 3 except the quantities are plotted as a function of the
    normalized absorption strength, averaged between the epochs being compared.}
\end{figure}
\begin{figure}
  \includegraphics[width=84mm]{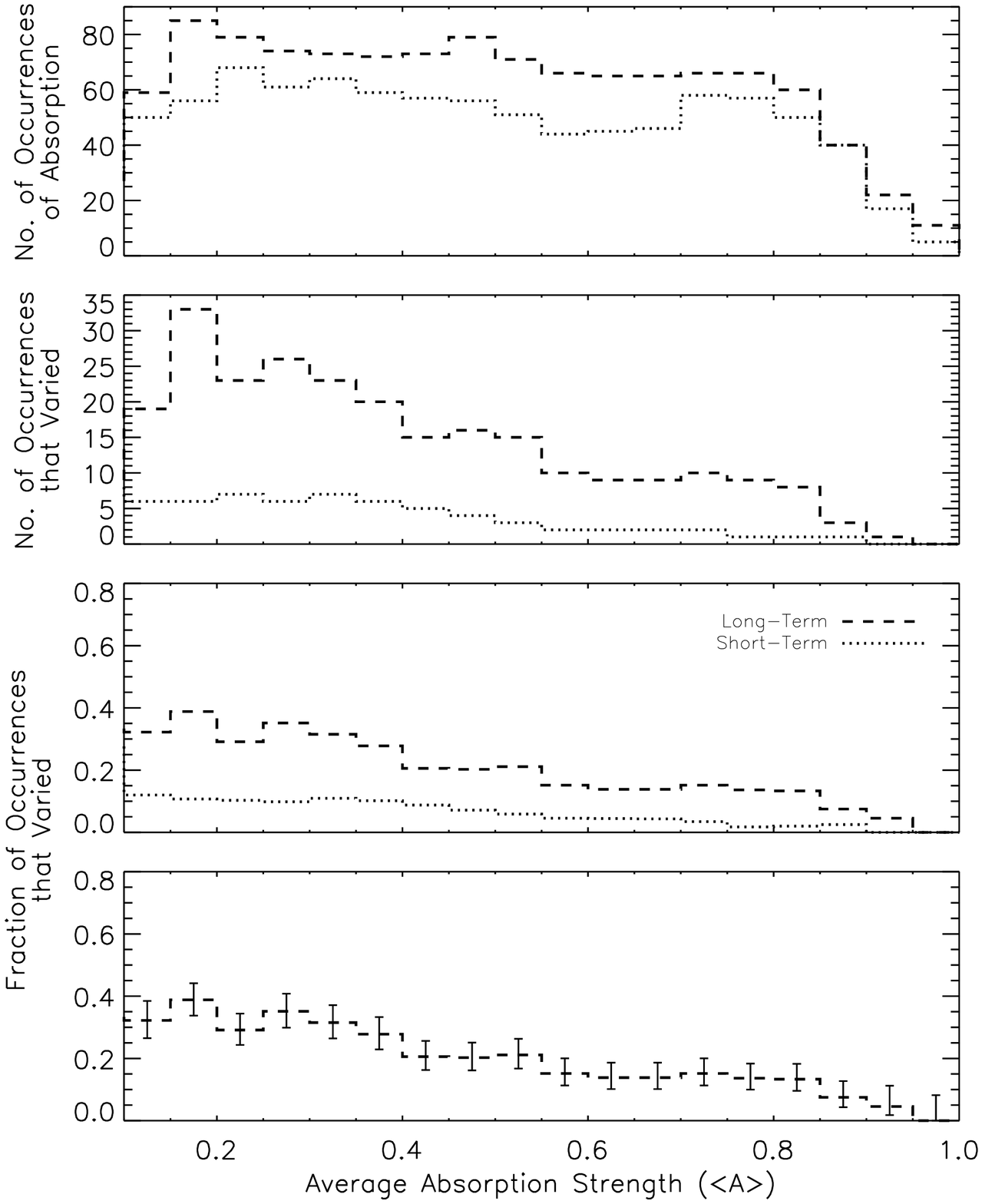}
  \caption{The top two panels show the total number of occurrences of absorption, and the
    number of those occurrences that varied, at each strength. The third panel is the second 
    panel divided by the top one. The bottom panel displays the long-term data with 1$\sigma$ 
    error bars.}
\end{figure}
Another factor that might affect the results in Figure 3 is that BALs at higher velocities tend to be weaker (Figure 2). In Figure 4, we examine \civ\ BAL variability as a function of absorption strength. Using the values of $\langle A\rangle$ calculated in \S3, and a bin size of 0.05, we count how many quasars have absorption in each bin. We then count how many quasars have at least one occurrence of variability in each absorption strength bin. Each quasar is counted at most once in each bin, even if it has the same absorption strength at more than one location in the spectrum. As in Figure 3, the third panel is the second panel divided by the top one, in this case yielding the fraction of quasars with a BAL that varied in each absorption strength bin. Figure 4 indicates that quasars are more likely to exhibit BAL variability at weaker absorption strengths. As in Figure 3, we calculate errors for each absorption strength value. We again perform a least-squares fit, which gives a slope of $-$0.487 $\pm$ 0.095 fraction per unit absorption strength. The slope is non-zero at a 5-sigma significance. We repeat the least-squares fit with only the bins with the most data, as for Figure 3; this restricts the fit to all of the bins except the one at $\langle A\rangle$ $>$ 0.95. The resulting slope is $-$0.476 $\pm$ 0.100 fraction per unit absorption strength, which is non-zero at a 4.8-sigma significance.

Figure 4, however, does not account for multiple occurrences of the same absorption strength within an individual quasar. In \S3.3, we define a method of dividing each BAL into $\sim$1200 \kms\ bins, and we now consider these bins as individual occurrences of absorption. In Figure 5, we plot the number of these occurrences at each absorption strength value in the top panel and then the number of these occurrences that varied in the second panel. The third panel is the second panel divided by the top one, showing the fraction of occurrences at each absorption strength that varied. An individual BAL may be counted more than once at each absorption strength value because we consider portions of each BAL trough in this figure. The bottom panel is the same as the third panel, except it shows just the long-term data with error bars, calculated as in Figures 3 and 4. A least-squares fit for the long-term data in Figure 5 gives a slope of $-$0.380 $\pm$ 0.052 fraction per unit absorption strength, which is non-zero at a 7-sigma significance. Figure 5 confirms that weaker portions of BAL troughs are more likely to vary than stronger ones.

\begin{figure}
  \includegraphics[width=84mm]{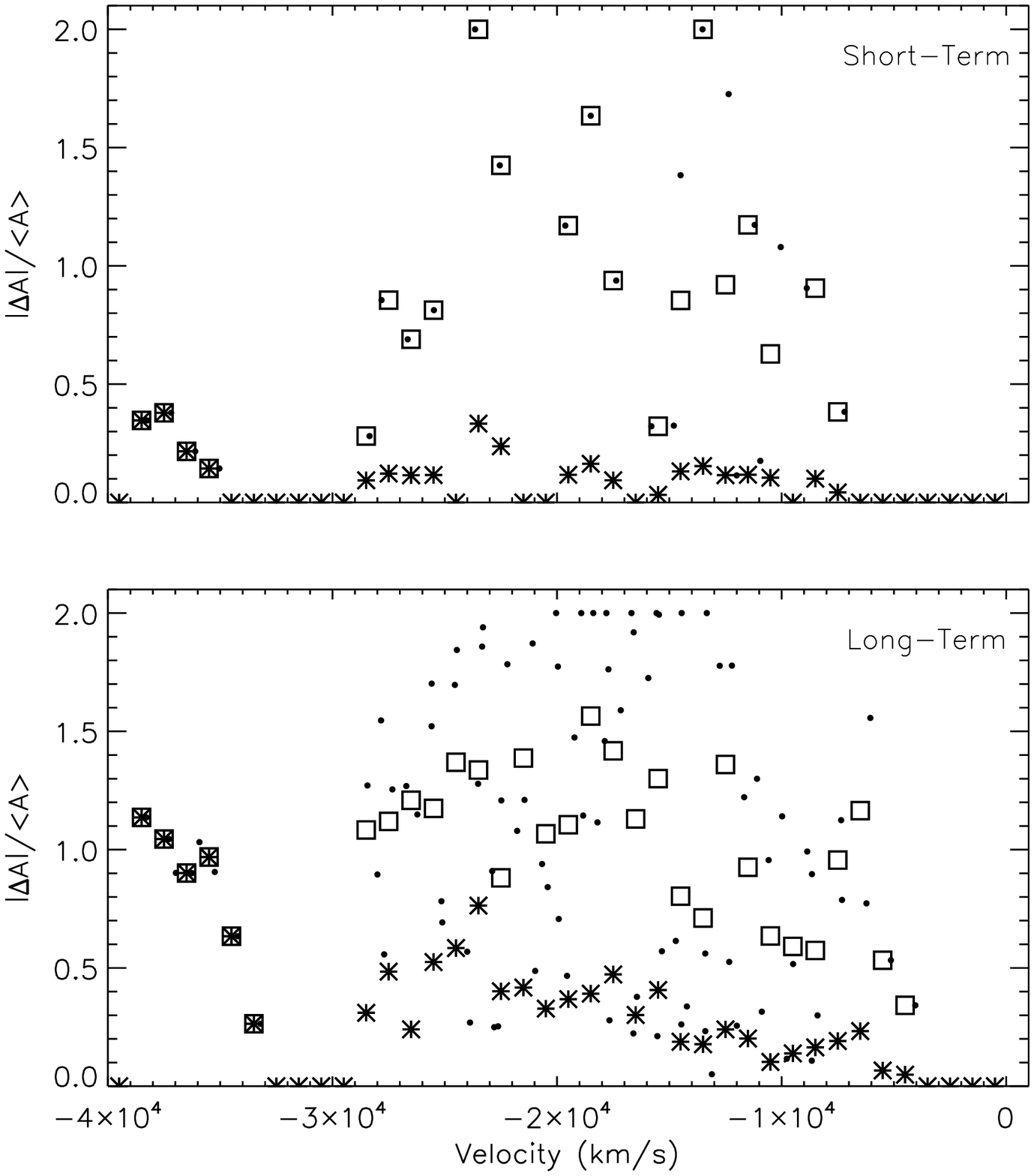}
  \caption{Fractional change in the depth of BAL absorption as a function of velocity. The
    symbols follow the same pattern as Figure 2, with individual points representing variability in a
    single quasar. Here, the open squares indicate the average of these variable, non-zero points
    in each velocity bin. The asterisks show the average if the non-varying absorption intervals
    are included. The top panel shows the short-term results, while the bottom panel plots
    long-term data. We note that the maximum value of $|\Delta A|/\langle A\rangle$ is 2 (see
    Figure 7), which indicates absorption appearing or disappearing completely.}
\end{figure}
Figures 3 to 5 compare the \emph{incidence} of variability to properties of the BALs. Next, we use the values of $|\Delta A|/\langle A\rangle$, calculated in \S3, to compare the \emph{amplitude} of variability to the same BAL properties.

In Figure 6, we plot this fractional change in absorption strength as a function of the outflow velocity, using bin sizes of 1000 \kms\ width, as in Figure 3. In each velocity interval, the individual points represent variability in a different quasar. The average values of $|\Delta A|/\langle A\rangle$ for just the variable (non-zero) points in each bin are shown by the open squares. The asterisks show the average if the non-varying absorption intervals are included. This plot does not indicate a statistically significant trend between fractional change in absorption strength and the outflow velocity.

\begin{figure}
  \includegraphics[width=84mm]{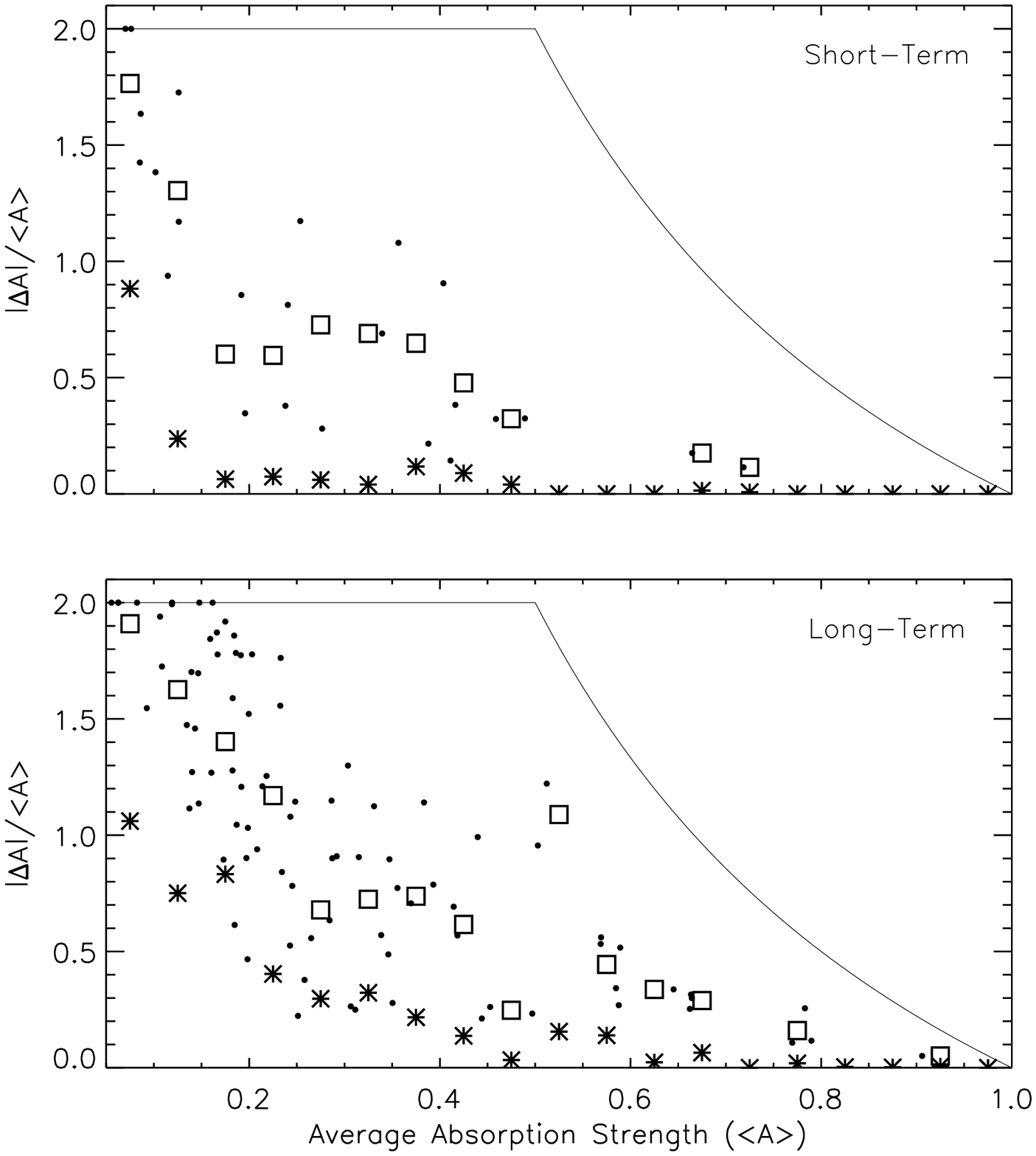}
  \caption{Same as Figure 6, except the fractional change is now plotted as a function of
    average normalized absorption strength. In this plot, individual quasars can contribute
    multiple times to the measurements at a given absorption strength. The solid curve
    represents the maximum possible value of $|\Delta A|/\langle A\rangle$.}
\end{figure}
In Figure 7, we plot the fractional change as a function of average normalized absorption
strength, similar to Figure 6. In this plot, however, individual quasars will contribute
multiple times to the measurements at a given absorption strength if the quasar has
variability in both the red and blue sides of an absorption trough and/or in multiple troughs. The individual points represent variability at different velocities in each quasar. The average of all the variable (non-zero) points in each bin are shown by the open squares, while the asterisks show the average $|\Delta A|/\langle A\rangle$ in each bin if non-varying absorption regions are included. This plot indicates that the fractional change in strength is greater in weaker features. However, the errors in the values of $|\Delta A|/\langle A\rangle$ are correlated with the values of $\langle A\rangle$. At an $\langle A\rangle$ of 0.15, the typical error values range from $\sim$0.11-0.33, whereas at $\langle A\rangle$ of 0.60, the error values are $\sim$0.02-0.06. A larger error at smaller values of $\langle A\rangle$ should cause a larger dispersion in the points in that region of Figure 7, but instead the values of $|\Delta A|/\langle A\rangle$ converge towards the maximum possible value of $|\Delta A|/\langle A\rangle$. The dispersion of points due to errors in $|\Delta A|/\langle A\rangle$ alone cannot explain the trend in this figure. Instead, this figure indicates that we see weaker absorbers, or weaker portions of BALs, appear or disappear completely, but strong features do not typically appear or disappear over the time-scales that our data cover.

Beyond looking at trends with velocity and absorption strength, it is instructive to examine
the relationship between these two parameters. Figure 2 displays $\langle A\rangle$ as a function of velocity, revealing that these quantities are correlated. The lack of a clear trend in Figure 6 is surprising given that weaker systems are more variable (Figure 7) and they tend to appear more often at higher velocities (Figure 2). Evidently, the correlation of $|\Delta A|/\langle A\rangle$ with absorption strength is stronger than the correlation with velocity. We also investigate how $|\Delta A|$ varies with velocity and with $\langle A\rangle$. We do not include the plots here, but we find no trend between $|\Delta A|$ and velocity or absorption strength.

We do not find any significant differences in the trends described above between the short-term and long-term data. The main difference is in the incidence of variability, 39\% in the short-term versus 65\% in the long-term data. The long-term data also show slightly larger changes in absorption strength, i.e., $|\Delta A|$, compared to the short-term data. The average value of $|\Delta A|$ is 0.16 $\pm$ 0.08 in the short-term and 0.22 $\pm$ 0.10 in the long-term.

We also note again that we include two quasars (0302+1705 and 0846+1540) from the Lick sample (\citealt{bar}) that are not technically BALs. However, these two objects do not significantly affect the results. If we omitted them, then the fraction of quasars with variability would be 35\% (6/17) in the short-term and 67\% (14/21) in the long-term. In the case of 0302+1705, it does not qualify as a BALQSO because its broad absorption feature is at low velocities (centered at $\sim$$-$2100 \kms). It also has $\langle A\rangle$ reaching $>$0.90. It is a strong feature at low velocity that does not vary, fitting the trends discussed above. In 0846+1540, there is broad absorption at low velocity (centered at $\sim$$-$2900 \kms), with average $\langle A\rangle$ of $\sim$0.27, and at high velocity (centered at $\sim$$-$26,800 \kms), with average $\langle A\rangle$ of $\sim$0.18. This also fits the overall trend for higher incidence of variability at higher velocities.

\section{Summary and Discussion}

We have analyzed short-term and long-term \civ\ $\lambda$1549 BAL variability in a sample of
24 quasars, and we looked for trends with the outflow velocity and absorption strength. We
found that 39\% (7/18) of the quasars varied in the short-term, intervals of 0.35 to 0.75
yr, whereas 65\% (15/23) varied in the long-term, intervals of 3.8 to 7.7 yr. Overall, 67\%
(16/24) of the quasars varied. We find a trend for variability to occur more often at higher
velocities (Figure 3) and in shallower absorption troughs (or shallower portions of absorption troughs; Figures 4 and 5). When looking at the fractional change in strength of the varying absorption features, there is no apparent significant correlation with velocity (Figure 6), but there is a trend for a larger fractional change in absorption strength in shallower features (Figure 7). We do include two objects with BI=0, but the variability in the broad absorption in these two quasars is consistent with the trends described above.

\citet{gibson08} studied \civ\ BAL variability on time-scales of $\Delta$t$\sim$3-6 yrs in 13
quasars with two epochs of data and found that 92\% (12/13) varied. This is larger than our
overall percentage of 67\%. This is surprising because the selection criteria adopted by
\citet{bar} for the sample used here might be biased toward more variable sources (\S2). Evidently, this bias is small or nonexistent. The difference between our result and \citet{gibson08} might be due to the small numbers of objects studied and/or the more conservative approach we took for identifying variable absorption.

\citet{L07}, which is a study of BAL variability on time-scales of $<$1 yr, find an inverse correlation of fractional change in equivalent width (EW) with average EW. While EW is a different calculation than our measure of absorption strength, $\langle A\rangle$, encapsulating the entire BAL profile in one number instead of considering separately the different profile regions (\S3.3), this result is consistent with our findings that fractional change in $\langle A\rangle$ correlates with $\langle A\rangle$. \citet{L07} also find no overall trend between fractional change in EW and outflow velocity, which is again consistent with our results. While not plotted here, we do not find a significant correlation between $|\Delta A|$ and velocity or absorption strength, which is consistent with the analysis of \citet{gibson08}. \citet{gibson08} also calculate the change in absorption strength, with values concentrated in the range 0.15-0.25. We find an average value of $|\Delta A|$ for the long-term data of 0.22 $\pm$ 0.10.

Both \citet{gibson08} and \citet{L07} comment on the lack of evidence for acceleration in the
BAL features they study. We also find no evidence for acceleration in our sample. The BAL
variations only involve the absorption trough growing deeper or becoming shallower.
We note, however, that limits on the acceleration are difficult to quantify for BALs
because, unlike narrower absorption lines, they do not usually have sharp features to
provide an accurate velocity reference. BALs also exhibit profile variability, which can
shift the line centroid without necessarily corresponding to a real shift in the velocity
of material in the outflow.

In this work, we compare two different time intervals, and we do not find a significant
difference between the short-term and long-term data in the trends described in \S4.
However, the incidence of variability and
the typical change in strength is greater in the long-term than in the short-term. This
indicates that BALs generally experience a gradual change in strength over multi-year time
scales, as opposed to many rapid changes on time-scales of less than a year. This will be
investigated more in a later paper where we will include all the epochs from our data set.

Variability has been investigated in other types of absorbers that are potentially related
to BALs. \citet{narayanan} and \citet{wise04} looked for variability ($\Delta$t$\sim$0.3-6
yrs) in low-velocity narrow absorption lines (NALs) and found that 25\% (2/8) and 21\%
(4/19), respectively, of the NALs they studied varied. {\citet{paola10} examined
variability in mini-BALs, which are absorption lines with widths larger than those of NALs
and smaller than those of BALs.  They observed \civ\ mini-BALs in 26 quasars on
time-scales of $\sim$1 to 3 years and found variability in 57\% of the sources. This number
is similar to our result for BALs, however this comparison is complicated by a differing
number of epochs and different time baselines.

\citet{paola10} found no correlation between incidence of mini-BAL variability and absorption strength or outflow velocity. However, this comparison might be skewed by the fact that mini-BALs tend to be weaker than BALs. For example, none of the mini-BALs studied by \citet{paola10} have depths $A>0.6$, while some of the BAL features in our sample have $A\sim1$. There are also differences in the velocity distributions. The mini-BAL distribution in \citet{paola10} peaks at $\sim$$-$20,000 \kms, whereas the distribution of BAL absorption in our data peaks at $-$15,000 to $-$12,000 \kms (Figure 3). More work is needed, e.g., with larger samples, to control for these differences in the absorption characteristics and compare BAL and mini-BAL variabilities on equal footing. Comparisons like that could provide valuable constraints on the physical similarities and relationship between BAL and mini-BAL outflows.

Two possibilities for the cause of BAL variability are the movement of gas across our line
of sight and changes in ionization.  We found a trend for lines at higher velocity to vary more often, which suggests movements of clouds since faster moving gas might vary more. Moreover, this
is a global trend, dependent on velocity in an absolute sense, rather than relative velocity
within a given trough. However, it is possible that the root cause of the variability is
associated with the strength of the lines, and weaker lines happen to appear at higher
velocities (see Figure 2 and \citealt{korista93}). In either case, faster moving
material tends to have lower optical depths or cover less of the continuum source. Measured
values of $\langle A\rangle$ thus provide either a direct measure of optical depth, or covering
fraction in the case of saturated absorption. With the current results, we cannot yet
disentangle the trends in velocity and $\langle A\rangle$ with certainty.

Evidence from BAL variability studies nonetheless supports an interpretation of varying
covering factor. In addition to greater variability at higher velocities, we found changes
over small portions of the BAL troughs, rather than entire BALs varying, which can be
understood in terms of movements of sub-structures in the flows. Changes in the ionizing
flux should globally change the ionization in the flow, and we would expect these changes to
be apparent over large portions of the BAL profiles and not in just some small velocity
range. \citet{fred08} looked at the variation in the \civ\ and \siiv\ BALs in one quasar and
found that the BALs were locked at roughly a constant \siiv/\civ\ strength ratio. They
attribute this result to the movements of clouds that are optically thick in both lines. If
the level of ionization is constant between observations or if $\tau\gg$ 1, then the changes
in absorption strength in our sample are directly indicative of changes in the covering
fraction. We will examine the \siiv/\civ\ line ratio in our BAL data in a subsequent paper.

On the other hand, studies of NALs support the possibility of changes in ionization causing the line variability. \citet{misawa07} and \citet{hamann10a} observed quasars where multiple NALs
varied in concert, which suggests global changes in the ionization of the outflowing gas. If
a change in ionization is causing the variability in the BAL outflows, then the optical
depth ($\tau$) of the BALs is changing. Features with lower optical depth are much more
sensitive to changes in ionization than those with higher optical depth. The outflowing gas
is presumably photoionized by the flux from the continuum source, and changes in the
continuum flux could cause a change in the ionization of the outflowing gas. However,
studies to date have not found a strong correlation between continuum variability and BAL
variability (\citealt{L07}, \citealt{gibson08}, \citealt{bar}, \citealt{barlow92}), casting
doubt on ionization changes causing BAL variability.

Finally, we use the bolometric luminosities of the quasars in our sample to derive characteristic time-scales of the flows for comparison to the observed time-scales. The bolometric luminosities for the quasars in our sample range from $\sim2\times 10^{46}$ to $3\times 10^{47}$ ergs s$^{-1}$, and the average is $\sim7\times 10^{46}$ ergs s$^{-1}$. These values are based on the observed flux at 1450 \AA\ (rest-frame) in absolute flux-calibrated spectra from \citet{bar}, a cosmology with $H_o = 71$ \kms\ Mpc, $\Omega_M = 0.3$, $\Omega_{\Lambda}=0.7$, and a standard bolometric correction factor $L\approx 4.4\lambda L_{\lambda}(1450 {\rm \AA})$. Based on the average value of $L \sim 7\times10^{46}$ ergs s$^{-1}$, a characteristic diameter for the continuum region at 1550 \AA\ is $D_{1550}\sim 0.006$ pc and for the \civ\ BEL region is $D_{\rm CIV}\sim 0.3$ pc, assuming $L = 1/3L_{edd}$ and $M_{BH} = 1.4\times10^{9}M_{\sun}$ (\citealt{peterson04}, \citealt{bentz07}, \citealt{gaskell08}, \citealt{hamann10b}). If the outflowing gas is located just beyond the \civ\ BEL region and it has a transverse speed equal to the Keplerian disk rotation speed, then the gas cloud would cross the entire continuum source in $\sim$1 yr. The typical time-scale in our short-term data is $\sim$0.5 yr and the typical change in absorption strength is at least 10\%, which nominally requires the moving clouds to cross at least 10\% of the continuum source. This implies transverse speeds $>$1000 \kms\ and thus radial distances that are conservatively $<$6 pc if the transverse speeds do not exceed the Keplerian speed. A typical time-scale in the long-term data is $\sim$6 yr, which is enough time to cross the continuum source $\sim$7 times (if the gas is located just beyond the \civ\ BEL region). This suggests a fairly homogeneous flow for at least the cases where there is no long-term variability.

To place better constraints on the outflow properties and understand the underlying cause(s)
of the line variabilities, our following papers will make more complete use of our
entire data sample. We will examine the ratio of \siiv\ to \civ\ absorption to gain better
insight into the cause(s) of the variability, and we will utilize more complete time-sampling
instead of just looking at two specific ranges of time intervals. We have also obtained more
data on BAL variability at the shortest time-scales, $\Delta$t $\sim$ 1 week to 1 month, in
order to determine whether there is a minimum time-scale over which variability occurs. This
will provide a better constraint on the location of the outflowing gas.

\section*{Acknowledgments}

We thank an anonymous referee for helpful comments on the manuscript. Funding for the SDSS and SDSS-II has been provided by the Alfred P. Sloan Foundation, the Participating Institutions, the National Science Foundation, the U.S. Department of Energy, the National Aeronautics and Space Administration, the Japanese Monbukagakusho, the Max Planck Society, and the Higher Education Funding Council for England. The SDSS Web Site is http://www.sdss.org/.

\bibliographystyle{mn2e}

\label{lastpage}

\end{document}